\documentclass[12pt,twocolumn]{openjournal}

\usepackage{xcolor}
\usepackage{textgreek}
\usepackage[utf8]{inputenc}
\usepackage[T1]{fontenc}
\usepackage[english]{babel}
\usepackage[hidelinks]{hyperref}
\hypersetup{
    colorlinks=true,
    linkcolor=blue,
    filecolor=blue,      
    urlcolor=blue,
    citecolor=blue,
}
\usepackage{color,colortbl}
\usepackage{tensind}
\tensordelimiter{?}
\DeclareGraphicsExtensions{.bmp,.png,.jpg,.pdf}
\usepackage{verbatim}
\usepackage{needspace}
\usepackage[normalem]{ulem}
\usepackage{orcidlink}
\usepackage{soul}
\usepackage{xspace}
\usepackage{booktabs}
\usepackage{float}

\usepackage{caption}
\newcommand{\reb}{{\sc \tt REBOUND}\xspace}

\newcommand{\airball}{{\sc \tt AIRBALL}\xspace}

\newcommand{\whfastAVX}{{\sc \tt WHFast512}\xspace}
\newcommand{\whckl}{{\sc \tt WHCKL}\xspace}
\newcommand{\ias}{{\sc \tt IAS15}\xspace}

\newcommand{\fig}[1]{Figure~\ref{#1}}
\newcommand{\eq}[1]{Eq.~\ref{#1}}

\begin{document}

\title{A substellar flyby that shaped the orbits of the giant planets}
\shorttitle{A substellar flyby that shaped the orbits of the giant planets}

\author{Garett Brown\orcidlink{0000-0002-9354-3551}}
\email{garett.brown@mail.utoronto.ca}
\affiliation{Dept. of Physical and Environmental Sciences, University of Toronto at Scarborough, Toronto, Ontario, M1C 1A4, Canada}
\affiliation{Dept. of Physics, University of Toronto, Toronto, Ontario, M5S 3H4, Canada}

\author{Renu Malhotra\orcidlink{0000-0002-1226-3305}}

\affiliation{Lunar and Planetary Laboratory, The University of Arizona, Tucson, Arizona, 85721-0092, USA}

\author{Hanno Rein\orcidlink{0000-0003-1927-731X}}

\affiliation{Dept. of Physical and Environmental Sciences, University of Toronto at Scarborough, Toronto, Ontario, M1C 1A4, Canada}
\affiliation{Dept. of Physics, University of Toronto, Toronto, Ontario, M5S 3H4, Canada}
\affiliation{Dept. of Astronomy and Astrophysics, University of Toronto, Toronto, Ontario, M5S 3H4, Canada}

\shortauthors{Brown et al.}

\begin{abstract}
The modestly eccentric and non-coplanar orbits of the giant planets pose a challenge to solar system formation theories which generally indicate that the giant planets emerged from the protoplanetary disk in nearly perfectly circular and coplanar orbits. 
We demonstrate that a single encounter with a 2--50 Jupiter-mass object, passing through the solar system at a perihelion distance less than 20 AU and a hyperbolic excess velocity of 1--3~$\mathrm{km\,s}^{-1}$, can excite the giant planets' eccentricities and mutual inclinations to values comparable to those observed. 
We describe a metric to evaluate how closely a simulated flyby system matches the eccentricity and inclination secular modes of the solar system.
We estimate that there is about a 1-in-9000 chance that such a flyby occurs during the solar system's residence in its primordial cluster and produces a dynamical architecture similar to that of the solar system.
The scenario of an ancient close encounter with a substellar object offers a plausible explanation for the origin of the moderate eccentricities and inclinations and the secular architecture of the planets.
We discuss some broader implications of disruptive flyby encounters on planetary systems in the Galaxy.
\end{abstract}


\maketitle


\section{Introduction}
\label{sec:intro}

Since antiquity, astronomers have sought to unravel the intricate patterns of planetary motion, refining their models over centuries to better align with empirical observations.
In some ways, the historical expectations that planets move in perfect circles align with modern planet formation theories which suggest that planets emerge from protoplanetary disks in nearly circular and coplanar orbits \citep{GoldreichTremaine1980, Artymowicz1992, Pollack96, LubowSeibertArtymowicz1999, PapaloizouNelsonMasset2001, IdaLin2004, KleyNelson2012, Armitage2020}.
Consequently, the origin of the modestly eccentric and modestly non-coplanar orbits of the giant planets of the solar system has been an abiding puzzle \citep{Tsiganis2005, Morbidelli2007, Nesvorny2012, Raymond2018, Nesvorny2018, Clement2021, Griveaud+2024}.
In this paper, we investigate external perturbations from flybys of stellar and substellar objects as a mechanism for the origins of the eccentricities and inclinations of the planets.

As background to our investigation, we note that the present-day eccentricities and inclinations of the solar system planets are the instantaneous manifestations of their long-term variations under mutual planetary perturbations.
In the linear approximation of the Laplace-Lagrange secular perturbation theory, these long-term (secular) variations of the eccentricities are separable from those of the inclinations, and both are approximated by the superposition of linear modes whose frequencies are determined by the planetary masses and semi-major axes \citep{BrouwerClemence1961, Murray1999, TremaineBook2023}.
Henceforth, we will refer to these linear modes as the fundamental secular modes.
The fundamental secular modes and their long-term variations have been the subject of many studies as they are crucial for understanding the long-term dynamical stability of the solar system \citep{Laskar1988, Laskar1990, Laskar1996, ItoTanikawa2002, LaskarGastineau2009, Zeebe2015, MogaveroLaskar2021, MogaveroLaskar2022, HoangMogaveroLaskar2022} as well as the climate cycles of Earth \citep[e.g.][]{Laskar+2004, Laskar2011}.
These studies show that the semi-major axes and the fundamental secular mode amplitudes of the planets are largely well preserved over billion-year timescales.
Assuming that the giant planets' masses and orbital separations were largely determined by formation processes in the primordial protoplanetary disk (with only minor changes subsequently), one infers that the amplitudes of the eccentricity and inclination secular modes must originate from some form of internal or external perturbation in the ancient solar system, after planet formation.
Possible internal perturbations include planet-planet interactions, either from scattering encounters or mean motion resonances; external perturbations include stellar flybys.

In our investigation of perturbations from flybys of stellar and substellar objects, we initially focus on the giant planets because their orbital evolution is nearly unaffected by the terrestrial planets \citep{Laskar1990}.
Later we also include the terrestrial planets in our analysis.
Previous studies to constrain possible signatures of stellar flyby events in the solar system's history have looked at the orbital configuration of the small bodies in the outer solar system (the Kuiper belt, Sednoids, and Oort Cloud) \citep{Levison2004, Morbidelli2004, Kenyon2004, Brasser2006, Adams2010, HuangGladman2024, PfalznerGovindPortegies-Zwart2024}, and the long-term stability of the solar system \citep{Li2015, LiMustillDavies2019, BrownRein2022, Raymond2024, KaibRaymond2025}. 
Indeed, significant effort has been undertaken to investigate the stability and dynamics of planetary systems in general within the context of stellar flyby encounters \citep[and references therein]{Laughlin1998, Laughlin2000, Malmberg2007b, Spurzem2009, Malmberg2011, Hao+2013, Cai2017, LiMustillDavies2020, SchoettlerOwen2024}.
However, none of these investigations consider both extremely close (periastron distance less than $20\,\mathrm{AU}$) and substellar-mass objects (masses less than $50\,\mathrm{M}_\mathrm{J}$) together in the Sun's primordial star cluster.

In this paper, we consider close encounters from a typical open star cluster environment using an initial mass function that includes sub-stellar objects with masses ranging between $10^{-3}\,\mathrm{M}_\odot$ and $10^{2}\,\mathrm{M}_\odot$.
We focus on encounters with perihelia less than $20\,\mathrm{AU}$ of the Sun.
Due to the closeness of the flybys, we run full $N$-body simulations rather than analytical estimates to accurately resolve the close encounters. 
In total, we simulate $5\cdot 10^4$~flybys with the solar system giant planets initially on circular, coplanar orbits at their current semi-major axes.

We first describe the methods and our initial conditions in detail in Section~\ref{sec:methods}.
We then present the results of our simulations, including our best matching flyby and the bigger statistical picture in Section~\ref{sec:results}. 
Finally, we detail our conclusions in Section~\ref{sec:conclusions}.

\section{Methods}
\label{sec:methods}

\subsection{Stellar Environment}
\label{sec:stellarenvironment}

We use a Monte Carlo approach to investigate the impact of various flybys on the solar system by simulating $5\cdot 10^4$ flyby objects from an open star cluster environment with a mass density of $100\,\mathrm{M_\odot\,pc^{-3}}$.
For each flyby, we sample a mass, velocity, and impact parameter along with a random orientation centered on the Sun \citep{Zink2020}.

The initial mass function (IMF) we use to generate the flyby masses $m_\star$ is the Chabrier IMF \citep{Chabrier2003} which smoothly combines a log-normal IMF for single stars with masses less than one solar mass and the standard power-law IMF \citep{Salpeter1955} for masses above one solar mass.
We sample masses from the IMF in the range from $10^{-3}\,\mathrm{M}_\odot$ to $10^2\,\mathrm{M}_\odot$.

We note that there is some uncertainty in the IMF on the low-mass end.
Nevertheless, IMFs seem to be in reasonable agreement across different environments and star forming regions as long as the conditions are not too extreme \citep{Bastian2010}.
\cite{Raghu2024} show that theoretical models of brown dwarf formation are consistent with the observed power-law of the mass probability distribution function (PDF) obtained by microlensing observations by \cite{ChabrierLenoble2023}.
Yet, \cite{Kirkpatrick2024} find that previous IMFs might over-estimated the abundance of some portions of the substellar mass regime.

\begin{figure}
    \centering
    \includegraphics[width=\columnwidth]{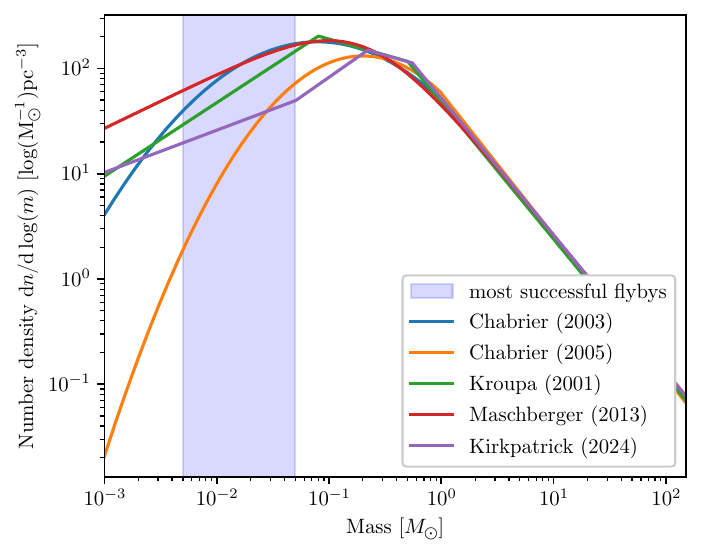}
    \caption{A comparison of IMFs. While there is uncertainty in the substellar regime, most IMF models are consistent in modeling the number density of substellar objects within one order of magnitude. 
    This paper uses the Chabrier (2003) IMF. 
    We also highlight the approximate mass range for which flybys successfully produce a dynamical architecture similar to the solar system (see also Fig.~\ref{fig:flyby-params}).
    }
    \label{fig:imfs}
\end{figure}

Because of this uncertainty we compare various popular IMFs \citep{Kroupa2001, Chabrier2003, Chabrier2005, Maschberger2013, Kirkpatrick2024} in \fig{fig:imfs}.
Each IMF is normalized assuming a mass density of $100\,\mathrm{M_\odot\,pc^{-3}}$.
For the IMFs shown, the number of objects per cubic parsec, $n$, varies from $181\,\mathrm{pc^{-3}}$ \citep{Chabrier2005} to $332\,\mathrm{pc^{-3}}$ \citep{Maschberger2013}.
Roughly half of the objects have substellar masses, $m < 80\,\mathrm{M_J}$, corresponding to an expected number density of substellar objects ranging from $52\,\mathrm{pc^{-3}}$ to $175\,\mathrm{pc^{-3}}$, respectively.

We sample hyperbolic excess velocities $v_\infty$ from a Maxwell-Boltzmann distribution with a velocity dispersion of $1\,\mathrm{km\,s}^{-1}$ \citep{Girard1989, Adams2010}.
We sample impact parameters $b_\star$ to uniformly cover the cross-sectional area out to $75\,\mathrm{AU}$.
Gravitational focusing \citep{BinneyTremaine2008} naturally gives a perihelion which is less than the impact parameter with the host star.
For our simulations, all perihelia come within $55\,\mathrm{AU}$ with $90\%$ coming within $20\,\mathrm{AU}$.
We sample the incident angles isotropically to allow for all possible encounter geometries.
We initialize the solar system with the four giant planets on circular and coplanar orbits at their current semi-major axes, and randomly initialize their true anomalies.

\subsection{Integration Methods}
We use the open-source $N$-body integrator \reb \citep{ReinLiu2012} along with \airball \citep{airball}, a python package for setting up and running simulations of flybys.
We use the \ias \citep{ReinSpiegel2015} integrator (with the improved criteria for the timestep presented by \citealt{PhamReinSpiegel2024}) to integrate the flyby in a heliocentric reference frame starting at a distance of $10^{5}\,\mathrm{AU}$ which is on the order of the Sun's current Hill sphere with the Galaxy \citep{PhamRein2024}.
When the flyby object has receded to a distance of $10^{5}\,\mathrm{AU}$, we remove it from the simulation, and then propagate the planetary system for another 20 Myr using the \whckl or \whfastAVX integrators \citep{ReinTamayoBrown2019,Javaheri+2023} depending on whether we are integrating the giant planets or all eight planets respectively.

For each flyby system that did not experience any planetary ejections, we then compute the power spectra of the complex eccentricities and complex inclinations of the planets and compare those to the solar system using the log-spectral distance metric (see below).

\subsection{Metric for secular architectures}
\label{sec:metric}

For the purpose of comparing the orbital architecture of two systems, some previous studies have used the angular momentum deficit (AMD), an approximately conserved quantity of the solar system  planets \citep{Laskar1997, Laskar2000, TurriniZinziBelinchon2020}.
Others have used a comparison of the instantaneous eccentricities and inclinations (or their minimum-to-maximum ranges) for each planet \citep{Marzari2002, Tsiganis2005, Morbidelli2007}.
At a more granular level, some studies have used the amplitudes of the fundamental secular modes for quantitative comparisons of the secular architectures of model solar systems \citep{Morbidelli2009,Brasser2009}.
In the context of planet-planet close encounters in the solar system, \cite{Nesvorny2012} describe a four-step procedure to measure the success of planet-planet interactions accounting for the secular architecture of the solar system. 
In the Appendix (A.2--A.4), we comment further on these previously used metrics compared to our adopted metric, described below.

For a more comprehensive comparison, we would like to measure the closeness of not only the fundamental secular modes but also their long-term variations, since the fundamental secular modes are merely the linearized approximation of an inherently non-linear system \citep[cf.][]{LithwickWu2014}.
All the above reasons motivate us to develop a new metric which measures the distance of a simulated system from the real solar system without making any initial assumptions regarding their dynamical structure or architecture. 

Let us start by defining the root-mean-square log-spectral distance, 
a measure of the distance between two Fourier spectra, often used in signal processing applications such as speech recognition
\citep{Gray1976, Rabiner1993}.
It compares the power spectra of two signals to evaluate the similarity of the amplitudes for each frequency as
\begin{equation}
    \mathcal{D}(X,\hat X) = \sqrt{\frac{1}{N}\sum_{k=1}^{N}\left(\log_{10} |X_k| - \log_{10} |\hat{X}_k|\right)^2}
    \label{eq:lsd}
\end{equation}
where $X_k$ and $\hat{X}_k$ are the complex-valued, discrete Fourier transforms (FFT) for the $k^\mathrm{th}$ frequency of the two signals being compared.
We make sure that the two signals are identical in length and sample spacing so that the amplitudes from the same frequencies are being compared.
This distance measure $\mathcal{D}$ also satisfies the properties of a metric, i.e. it satisfies the triangle inequality, is symmetric, and when $\mathcal{D} = 0$ there is a perfect match between the power spectra.
This can also be interpreted such that if the average ratio between the two spectra is $10^p$, i.e. $\hat X = 10^p X$, then $\mathcal{D}(X, \hat X) = \mathcal{D}(X, 10^p\,X) = \mathcal{D}(X, 10^{-p}\,X) = p$.

When we apply this distance measure to build a metric we equally consider it using the time evolution of both the complex eccentricities and inclinations of two systems.
The complex eccentricities and inclinations of a secular system for a given planet $j$ are defined as \citep{Laskar1990}
\begin{equation}
    z_j = e_j \exp\left(\mathtt{i}\varpi_j\right)
    \label{eq:complex_e}
\end{equation}
and
\begin{equation}
    \zeta_j = \sin(i_j/2) \exp\left(\mathtt{i}\Omega_j\right)
    \label{eq:complex_i}
\end{equation}
where $e_j$ is the eccentricity, $i_j$ is the inclination with respect to the invariant plane, $\Omega_j$ is the longitude of the ascending node,  $\varpi_j = \omega_j + \Omega_j$ is the longitude of the periapsis, $\omega_j$ is the argument of the periapsis, and $\mathtt{i} = \sqrt{-1}$.
We sample these values on evenly spaced time intervals over 20~Myr integrations with $N=2^{11}$ samples to then process with an FFT.
We use $\mathcal{D}$ in \eq{eq:lsd} to build a metric that evenly weighs the signals of the complex eccentricities and complex inclinations for each planet.
We define our metric as
\begin{equation}
    \mathcal{M} = \frac{1}{2n}\sum_{j=1}^{n}\left(\mathcal{D}(\mathcal{F}[z_j],\mathcal{F}[\hat{z}_j]) + \mathcal{D}(\mathcal{F}[\zeta_j],\mathcal{F}[\hat{\zeta}_j])\right)\,,
    \label{eq:metric}
\end{equation}
comparing one system (identified using a hat) with $n$ planets to another system with $n$ planets, where $\mathcal{F}[\cdot]$ is the discrete Fourier transform.
Thus, $\mathcal{M}$ is the averaged values given by $\mathcal{D}$ over the FFTs of the complex eccentricities and inclinations of all the planets. 

\subsection{Reference Set of Solar Systems}
\label{sec:referencesystems}
In order to make a fair comparison between the flyby systems and the observed solar system, we use a reference set of solar system realizations from an ensemble of integrations \citep{BrownRein2020}. 
This ensures that our reference set representing the solar system's secular architecture samples not just one single 20~Myr realization but also samples the effect of uncertainties of the initial conditions and of the small changes in the solar system's power spectrum over long times, both of which arise from nonlinear effects.

We start with the nominal initial conditions on 1 January 2000, obtained from NASA's Jet Propulsion Laboratory (JPL) Horizons data service.
Our solar system ensemble uses these nominal initial conditions that are identical in every way except for a small variation ($< 0.5\,\mathrm{m}$) in the semi-major axis of Mercury from the present-day value (see \citealt{BrownRein2020} for more details).
We use all $96$ of the 5~Gyr solar system simulations and divide them into 20 Myr non-overlapping segments from which we sample the complex eccentricities and inclinations to process with our metric.
We note that the fundamental secular mode frequencies of the giant planets change by at most $15\,\mathrm{mas\,yr}^{-1}$ and their amplitudes vary by at most $9.5\cdot10^{-4}$, but that the power at other frequencies is more variable amongst the 20 Myr snapshots of the solar system taken over billions of years (see \fig{fig:power-spectra}).

Using the metric $\mathcal{M}$ and this set of 20 Myr segments from our reference solar system ensemble, we cross compare each segment of a random subset of $2500$ segments against another randomly selected subset of $2500$ segments to compute the distribution of values.
This distribution of values of $\mathcal{M}$ quantifies how similar the solar system is to itself at different times and with slightly different initial conditions (within observational uncertainties).
The range of values given by $\mathcal{M}$ for this cross comparison of the solar system are between $0$ and $0.434$.

We want to determine the similarity of each post-flyby system to the solar system. 
To do that, we compare each post-flyby system to every 20~Myr segment of our full solar system ensemble.
This yields a distribution of values of the metric $\mathcal{M}$ for the simulated systems compared to the solar system. 
We can then compare this distribution to the distribution we obtain for the solar system compared to itself.

We found that this procedure provides a clear way of evaluating how close two systems are to each other and, most importantly, allows us to sort and rank them in closeness.
Moreover, this procedure avoids over-fitting to a specific manifestation of the solar system's long and diverse dynamical evolution.
What is not immediately clear is how to define the cutoff value of $\mathcal{M}$ that determines if a given system is considered to be a \emph{close match} to the solar system or not. 
For the present study, we define any flyby system with a metric value that overlaps with the reference solar system distribution as a \emph{close match} to the solar system.
We acknowledge that this choice is somewhat arbitrary and comes with several caveats. 
Note that the maximum value of $\mathcal{M}$ could be limited by the ensemble size.
Our ensemble size is large, $2500^2 > 6\cdot10^6$, so we don't expect this to be important in practice. 
Nevertheless, there exists the rare possibility that a solar system trajectory will eventually go unstable, for example with a Mercury--Venus collision \citep{Laskar+2004, LaskarGastineau2009}.
The closer we get to such an instability, the more the trajectory diverges from what is a more `typical' realization of the solar system.
Such a divergent trajectory will cause the range of metric values to increase significantly above the typical range and thus lead to more \emph{closely matching} flyby simulations.
However, if there aren't any significantly divergent trajectories (as is the case with our solar system ensemble, see \citealt{BrownRein2020}), then to use anything less than the entire range of the metric would be to claim that the solar system is not a close match to itself.
In summary, although it may be possible to find a better metric, the definition of what is considered a \emph{close match} to the solar system will likely always be subject to some reasonable but arbitrary quantitative choices.

\section{Results and Discussion}
\label{sec:results}

\subsection{Close Matches}

For each flyby simulation, we calculate the power spectrum of the complex eccentricities and inclinations over 20~Myr and compute its value of $\mathcal{M}$ relative to all the solar system power spectra.
Of the $5\cdot 10^4$ simulated flybys, we found $422$ cases ($0.844\%$) that lie within the range of the solar system's own $\mathcal{M}$ values.
As we described above, we consider these to be a \emph{close match} to the orbital architecture of our solar system.

Given that we initialize the pre-flyby circular orbits of the planets at their currently observed osculating semi-major axes, it is unsurprising that the fundamental secular modes of the solar system are prominent in the power spectra of the subset of flyby systems that incur only small changes in the planets' semi-major axes. In such cases, the fundamental secular mode \emph{frequencies} remain near the solar system values. 
However, the amplitudes of the fundamental modes as well as the power in the rest of the Fourier spectrum can vary considerably, depending on the flyby parameters. 
With the metric $\mathcal{M}$ we are demanding closeness of the entire power spectrum (of the complex eccentricity and complex inclination), not only closeness of the power in the fundamental secular modes. 
(In Appendix A.4, we describe a flyby case whose fundamental secular modes are closely matched but the full power spectrum is qualitatively and quantitatively very different from that of the solar system.)
This closeness of the entire power spectrum is realized in nearly $1\%$ of the flyby cases.
In the $\sim99\%$ of cases that are not close matches we find that the eccentricity modes are significantly under- or over-excited or at least one planet gets ejected in the first 20~Myr.

\begin{figure}
    \centering
    \includegraphics[width=\columnwidth]{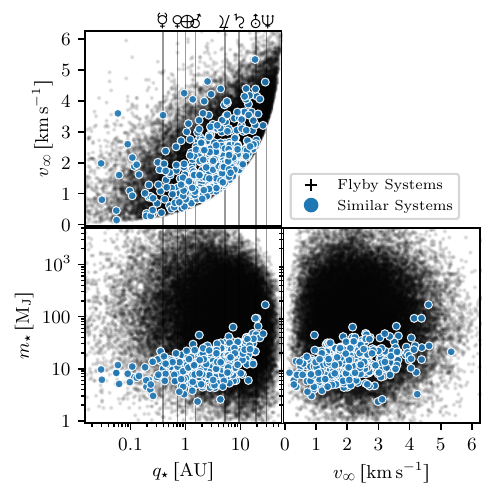}
    \caption{Here we show the distributions of the flyby parameters ($m_\star$, $q_\star$, $v_\infty$) in our simulations as well as the subset of those parameters that led to a \emph{close match} with the solar system.
    We highlight in blue the flyby systems that are similar to the solar system when compared to itself.
    \label{fig:flyby-params}
    }
\end{figure}

\fig{fig:flyby-params} shows the encounter parameters of all the flyby objects drawn from our stellar cluster parameters, highlighting in blue the \emph{close matches} to the solar system.
(See also \fig{fig:all-flyby-params} in the Appendix which shows all flyby parameters and their joint distributions in a corner plot.)
From here, we can see that the \emph{closest matching} flybys are those with substellar masses ($m_\star < 80\,\mathrm{M}_\mathrm{J}$), pass between the orbits of Mercury and Uranus ($\sim 0.4\,\mathrm{AU} < q_\star < 20\,\mathrm{AU}$ from the Sun), and have a hyperbolic excess velocity $v_\infty$ of 1--3~$\mathrm{km\,s}^{-1}$.
For typical velocities within an open star cluster, extremely close flybys ($q_\star < 50\,\mathrm{AU}$) with stellar masses ($m_\star > 80\,\mathrm{M}_\mathrm{J}$) almost always result in ejected planets or over-excited secular modes.
About $60\%$ of our $5\cdot 10^4$ simulations had at least one ejected planet in the first $20$ Myr.
Stellar mass flybys at greater distances are much less likely to eject a planet, but are also unable to sufficiently excite the eccentricity modes, particularly Jupiter's eccentricity \citep{Nesvorny2012, Griveaud+2024, BrownRein2022}.
It may be possible for a stellar mass encounter to enable the same degree of similarity between an initially circular, coplanar system and the solar system.
However, such a stellar flyby would need to be moving significantly faster than what is typically found within an open star cluster system, while also passing extremely close, making such cases exceedingly low probability.

The statistics of the flyby simulation outcomes are summarized as follows.
The most probable outcome is a disruption of the planetary system ($\sim60\%$ of cases).
In the next most probable outcome ($\sim39\%$ of cases) the eccentricities are with roughly equal probability either over or under excited, while the inclinations are consistently under excited when compared to the solar system.
In the remaining $0.844\%$ of cases, the outcome is a system similar to the solar system (as per our metric $\mathcal{M}$, \eq{eq:metric}).

\subsection{Best Match}
\label{sec:bestmatch}

\begin{figure}
    \centering
    \includegraphics[width=\columnwidth]{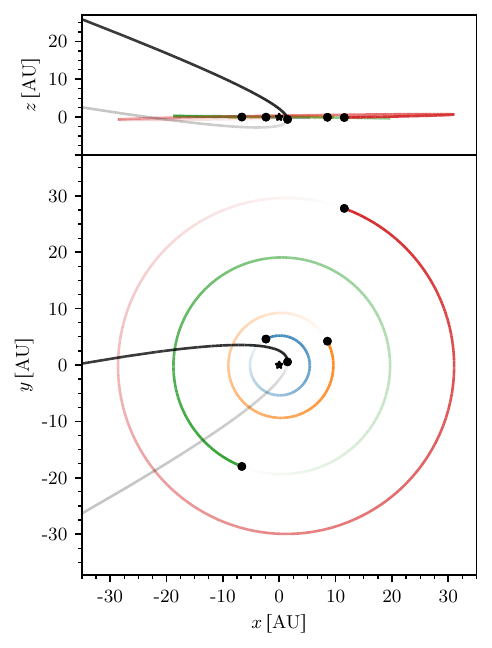}
    \caption{A snapshot of the flyby simulation that produces the best match to the solar system. The flyby parameters for the encounter are $m_\star = 8.27\,\mathrm{M}_\mathrm{J}$, $q_\star = 1.69\,\mathrm{AU}$, $v_\infty = 2.69\,\mathrm{km\,s}^{-1}$, and $i_\star = 131^{\circ}$.
    \label{fig:orbit-plot}}
\end{figure}

\begin{figure}
    \centering
    \includegraphics[width=\columnwidth]{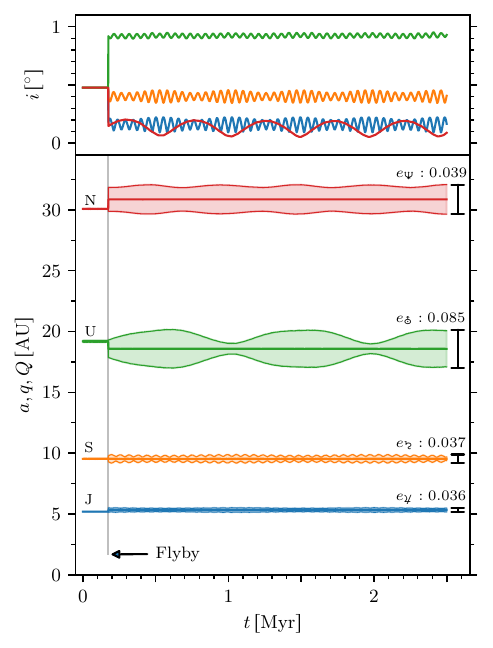}
    \caption{The time evolution of the giant planets before, during, and after the flyby shown in Figure~\ref{fig:orbit-plot}.
    The encounter time is indicated by the vertical line  at $t\approx0.2\,\mathrm{Myr}$.
    The upper panel shows the inclination of the planets (with respect to the final invariant plane) while the lower panel shows the semi-major axes, the perihelia, aphelia distances.
    The maximum eccentricity range of each planet is also indicated. \label{fig:flyby-effect}
    }
\end{figure}

\fig{fig:orbit-plot} visualizes the encounter that produces the best match to the solar system's secular architecture according to our metric.
In this case, the flyby object has a mass of $m_\star = 8.27\,\mathrm{M}_\mathrm{J}$, a perihelion of $q_\star = 1.69\,\mathrm{AU}$, a hyperbolic excess velocity of $v_\infty = 2.69\,\mathrm{km\,s}^{-1}$, and an inclination of $i_\star = 131^{\circ}$ with respect to the initial common plane of the planets. 
\fig{fig:flyby-effect} shows the time evolution of the semi-major axes $a$ of the giant planets along with the perihelia $q$, aphelia $Q$, and inclinations $i$ of the giant planets before, during, and after the flyby encounter.
The maximum eccentricities for each planet are indicated on the right side of the figure.

\begin{figure*}
    \centering
    \includegraphics[width=\textwidth]{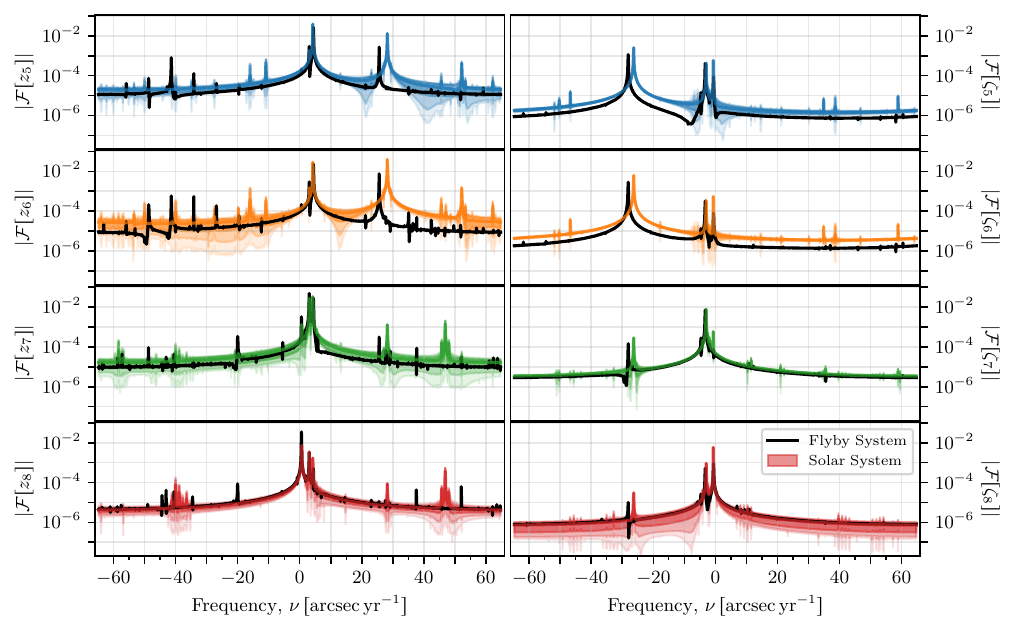}
    \caption{Power spectra of the complex eccentricities (left column) and complex inclinations (right column).
    The black line shows the best matching flyby case (as in Figures~\ref{fig:orbit-plot}~and~\ref{fig:flyby-effect}), while the color-shaded plots show the ensemble of spectra of the solar system taken in 20 Myr--long segments from an ensemble of 5 Gyr--long integrations. 
    The median power at each frequency of the solar system ensemble is shown as a dark coloured line, with lighter shaded regions representing the one--, two-- and three--$\sigma$ deviations. 
    The median power spectrum of the solar system is very similar to the simulated best case flyby.
    \label{fig:power-spectra}
    }
\end{figure*}

\fig{fig:power-spectra} shows the power spectra from the FFTs of the complex eccentricities and complex inclinations for the best matching flyby system (black line) overlaid with the ensemble of power spectra of the solar system.
In order to see how a particular flyby system compares to the spectral variation present in the true solar system, we determine the spread in the power for each frequency of the FFTs using the reference set of solar system integrations marking the median and the first three standard deviations.
The most prominent peaks in these spectra correspond to the frequencies of the fundamental secular modes \citep{Laskar1988, Laskar1990, Laskar1993, Sidlichovsky1996, Laskar2003}.
One can see that the amplitudes of the fundamental secular modes of the giant planets have been excited with a single flyby event to values similar to those of the real solar system.

\begingroup
\setlength{\tabcolsep}{10pt}
\begin{table}
    \centering
    \begin{tabular}{c|cc|cc}
      \toprule
      \midrule
      Mode & \multicolumn{2}{c|}{Freq.  $\mathrm{[arcsec\,yr^{-1}]}$} & \multicolumn{2}{c}{Amp. $|A|\cdot 10^3$} \\
      $\,$ & SS & Flyby & SS & Flyby \\
      \midrule
      $g_5$ & $4.26$ & $4.44$ & $44.17$ & $28.76$ \\
      $g_6$ & $28.25$ & $25.62$ & $48.21$ & $7.91$ \\
      $g_7$ & $3.09$ & $3.08$ & $29.03$ & $54.31$ \\
      $g_8$ & $0.67$ & $0.56$ & $9.14$ & $35.11$ \\
      \midrule
      $s_5$ & $--$ & $--$ & $--$ & $--$ \\
      $s_6$ & $-26.35$ & $-28.12$ & $7.85$ & $3.50$ \\
      $s_7$ & $-2.99$ & $-3.31$ & $8.87$ & $8.02$ \\
      $s_8$ & $-0.69$ & $-0.61$ & $5.81$ & $1.20$ \\
      \bottomrule
   \end{tabular}
   \caption{A comparison of the secular frequencies and amplitudes of the giant planets (SS) and our best matching flyby system that experienced a $m_\star = 8.27\,\mathrm{M}_\mathrm{J}$ encounter at $q_\star = 1.69\,\mathrm{AU}$, with $v_\infty = 2.69\,\mathrm{km\,s}^{-1}$. By convention, $s_5$ is the null frequency when inclinations are designated with respect to the invariant plane.
   \label{tab:outer}
   }

\end{table}
\endgroup

Table~\ref{tab:outer} provides a quantitative comparison between the present-day fundamental secular modes of the giant planets and the best matching flyby simulation.
For the four giant planets, there are four fundamental mode frequencies in the complex eccentricities, traditionally denoted as $g_5,g_6,g_7,g_8$, and three in the complex inclinations, $s_6,s_7,s_8$ \citep{BrouwerClemence1961}.
Table~\ref{tab:outer} shows that all the fundamental secular mode frequencies of the flyby system are a close match with those of the solar system.
As noted previously, this follows from the circumstance that the best matching flyby only mildly disturbs the semi-major axes of the planets: by $+2.6\%$, $-0.03\%$, $-3.2\%$, and $+2.6\%$ for Jupiter, Saturn, Uranus, and Neptune respectively.
(Recall that the fundamental mode frequencies depend only upon the semi-major axes and masses.)
We also see in Table~\ref{tab:outer} that the mode amplitudes of the best matching flyby system have magnitudes of a few hundredths in the eccentricity modes and a few thousandths in the inclination modes.
While similar in magnitude to those of the solar system, they differ in detail. 
The amplitudes of the inclination modes are a closer match than those of the eccentricity modes.

In particular, it has been recognized for some time that it is most challenging to excite the amplitude of the $g_5$ mode to values similar to the observed value by any of a number of plausible mechanisms considered in previous studies \citep{Nesvorny2012, Griveaud+2024}.
This is because the $g_5$ mode dominates Jupiter's orbit. As Jupiter has the largest planetary mass in the solar system, it is consequently the most resistant to orbital perturbations.
In that light, note that in the best matching flyby case, we find that the amplitude of the $g_5$ mode is about $65\%$ that of the present-day solar system.
Overall, we find that in $40\%$ of the $422$ \emph{closely matching} flyby systems, the amplitude of the $g_5$ mode is within a factor of two of the solar system's.
This is a notable success of substellar flybys as a mechanism to explain the secular mode amplitudes.

For the $g_6$ mode, there is mismatch in Table~\ref{tab:outer} as the amplitude of this mode in the best matching flyby simulation is much smaller than in the present-day solar system. We do not consider this significant because our simulations include other \emph{close matching} flyby systems that are more similar in the $g_6$ mode amplitude than our best matching case (though less similar in other ways), while still being close matches with the range of the solar system's power spectrum over billion-year timescales.

\needspace{2 \baselineskip}
\subsection{Inner Solar System}

Because many of the \emph{closely matching} systems experience flybys at perihelion distances in the region of the terrestrial planets, we simulate an additional $10^4$ circular, coplanar systems.
We use the exact same initial conditions for the giant planets and the exact same flyby event as shown in Figs.~\ref{fig:orbit-plot}~and~\ref{fig:flyby-effect}, but include the terrestrial planets at their current semi-major axes, also on circular, coplanar orbits but with random phases, as well as the effect of general relativity \citep{BrownRein2023}.

Remarkably, in the majority of cases we roughly reproduce the secular modes and amplitudes of the entire solar system.
In these simulations, we find that within the first 20 Myr following the flyby encounter, only $2\%$ of cases result in an ejected planet, with $4\%$ having at least one planet that reaches an eccentricity greater than $0.6$.
However, it becomes more difficult to make a specific quantitative claim of similarity for the following reasons.

The inclusion of the terrestrial planets means that there are many more degrees of freedom, and we find that the power spectra in these systems are noisier than the power spectra of the giant planets alone. 
Consequently, the distribution of the metric values $\mathcal{M}$ has a much larger range.
For the solar system compared to itself, the distribution of $\mathcal{M}$ is broadened, shifted, and has a more extended tail relative to its distribution when the inner planets are not included. 
(The reader is referred to the discussion at the end of Section~\ref{sec:referencesystems} for more context.) 
Due to the broader distributions of $\mathcal{M}$, many more of our flyby simulations qualify as close matches to the solar system.

\subsection{Flyby Probabilities}
\label{sec:probabilities}
The solar system is thought to have formed in an open star cluster that dispersed over a timescale of $10$--$100\,\mathrm{Myr}$ \citep{vandenBergh1981, ElmegreenClemens1985, BattinelliCapuzzo-Dolcetta1991, AdamsMyers2001, LadaLada2003, Adams2010, Pfalzner2013}.
Within the lifetime of such a cluster, we can estimate the flyby encounter rate as  the average of the product of the spatial number density $n$, the average relative velocity $v$, and the cross-sectional area $\sigma$ \citep{Adams2010}, $\Gamma = \langle n v \sigma \rangle$.
Assuming these variables are uncorrelated, the typical approach is to approximate the average of the product with the product of the averages.
With our adopted parameters (see Section~\ref{sec:stellarenvironment}) of $n=308\,\mathrm{pc}^{-3}$, $v = 1\,\mathrm{km/s}$, and $\sigma = \pi (75\,\mathrm{AU})^2$, we estimate the encounter rate as
\begin{equation}
    \Gamma \approx 1.3\cdot 10^{-4}\,\mathrm{Myr}^{-1}.
    \label{eq:encounter-rate}
\end{equation}
This encounter rate implies a 1-in-76 chance of a close flyby within $100$ Myr.
With the other IMFs discussed in Section~\ref{sec:methods}, we find similar probabilities of a close flyby over $100$ Myr: in the range from 1-in-130 \citep{Chabrier2005} to 1-in-71 \citep{Maschberger2013}.

This probability for close encounters may appear surprisingly high compared with previous works that have investigated stellar flybys of the solar system \citep{Jimenez-Torres2013, Pfalzner+2018, PfalznerVincke2020, Batygin+2020, BrownRein2022}.
The reason for this higher probability is the large population of substellar and super-Jovian mass objects included in the IMFs we have considered, whereas previous studies were limited to the stellar IMFs\footnote{\cite{Li2025} did look at planetary mass encounters in the context of Hilda asteroids.}.

As discussed above, our simulations show that, starting with initially coplanar and circular orbits, $0.844\%$ of such close flybys result in power spectra of the planets' eccentricities and inclinations similar to those of the real solar system.
Thus, the overall chance of the solar system planets acquiring their orbital eccentricities and inclinations via a close encounter with a substellar object is 1-in-9005.
For the other IMFs we considered, the chance is between 1-in-8412 and 1-in-15403.
The probabilities can also be  scaled to other stellar environments.
For example, if one chooses the lower end of the cluster dispersal timescale of 10~Myr, then the chance decreases by a factor of 10 because we make the simplifying assumtion that the stellar density remains static over the lifetime of the cluster.

\section{Conclusions}
\label{sec:conclusions}

In this paper, we explore a new scenario for the origin of the secular mode amplitudes of the giant planets.
We show that a single substellar object making a hyperbolic close encounter with the ancient solar system can account for the observed amplitudes in both the eccentricity and inclination fundamental secular modes.
We find the most favorable encounter to be a substellar object with encounter parameters $m_\star = 8.27\,\mathrm{M}_\mathrm{J}$, $q_\star = 1.69\,\mathrm{AU}$, $v_\infty = 2.69\,\mathrm{km\,s}^{-1}$, and $i_\star = 131^{\circ}$.
Our numerical simulations indicate that there is a 1-in-9005 chance that the necessary encounter parameters can be realized with random encounters within an open star cluster having properties expected for where the solar system formed.
Given that the estimated population of Sun-like stars in the Galaxy is on the order of $10^{10}$ and that stars are commonly formed in open star clusters, the $10^{-4}$ probability is \emph{not} negligible.
Even if we choose the most pessimistic scenario, with an IMF that has a much lower fraction of substellar objects \citep{Chabrier2005}, and a 10 times faster open cluster dispersal timescale, the probability is still~$7\cdot 10^{-6}$.
In other words, we don't need to look for a needle in a haystack to find a suitable encounter.
Additional numerical simulations also demonstrate that the terrestrial planets have a high probability of not only surviving such an encounter but also themselves acquiring secular mode amplitudes similar to those observed today.

The novelty of this work is two fold.
First, we invoke external perturbations to account for the solar system's secular modes; this is in contrast to invoking internal, planet-planet interactions of the giant planets \citep{Tsiganis2005, Morbidelli2007, Nesvorny2012, Raymond2018, Nesvorny2018, Clement2021, Griveaud+2024}.
Secondly, we include substellar mass objects (down to Jupiter mass) and extremely close flyby distances in the Sun's primordial star cluster in our simulations. 
This represents parameter regimes that have not been considered in previous studies of flyby encounters of the ancient solar system \citep{Levison2004, Morbidelli2004, Kenyon2004, Brasser2006, Adams2010, HuangGladman2024, PfalznerGovindPortegies-Zwart2024, Li2015, LiMustillDavies2019, BrownRein2022, Raymond2024}.

Note that there is a very large parameter space of initial conditions to explore in this scenario (velocity, direction, and mass of the flyby object, as well as the orbital phases of all planets) and the outcomes are sensitive to these initial conditions.
That we found a substantial fraction of \emph{close matches} to the solar system shows that this scenario is not only possible but surprisingly efficient.
This result holds regardless of the specific measure of closeness adopted (see Appendix~\ref{sec:success-criteria}--\ref{sec:oe} for more details).
However, let us acknowledge that identifying with very high precision the one specific set of flyby encounter parameters that reproduces every observed dynamical feature of the solar system is not possible---it would be as unique as the solar system itself.

Nevertheless, further exploration of this scenario is warranted.
For example, simulations could include a wider range of initial semi-major axes.
This would allow us to connect the present scenario to other earlier phases in planet formation theories including planet-disk interactions. 
Future investigations might also look into a more detailed study of the effects on the terrestrial planets, as well as the effect of substellar flybys on the dynamical excitation of minor planets in the asteroid belt and the trans-Neptunian belts, and the capture probability of irregular satellites by the giant planets.
Our results suggest that substellar flyby encounters should be taken into account when studying the formation and evolution of the Kuiper belt, trans-Neptunian objects (TNOs), and the Sednoids \citep{PfalznerGovindPortegies-Zwart2024}. 

Whereas this study focused on finding \emph{close matches} for our own solar system, a statistical analysis of all flyby outcomes might also reveal new insights into the formation history of extrasolar planetary systems which show different architectures compared to the solar system \citep{WinnFabrycky2015, Kane2024}.
To illustrate, we comment on some broader implications of the system disruption statistics in our flyby simulations.
We found that $\sim60\%$ of flyby cases are disruptive, resulting in ejected planets.
Combining this result with the estimated flyby encounter rate (\eq{eq:encounter-rate}), the implication is that a multi-planet system has a 1-in-1280 to 1-in-128 chance of a disruptive flyby over the 10--100 Myr cluster lifetime.
This estimate indicates that most planetary systems would avoid disruptive flyby encounters, roughly consistent with previous studies \citep[e.g.][]{Malmberg2011}.
We can also estimate that, for the $\mathcal{O}(10^{10})$ Sun-like stars in the Galaxy, the rate of planet ejection by flybys implies $\mathcal{O}(10^7-10^8)$ ejected planets contributing to the Galaxy's free-floating planet population.
These estimates must be taken with caution as they assume that the flyby outcomes have similar statistics for multi-planetary systems with varied planetary masses, initial orbital radii and open cluster properties.

 Lastly, let us also acknowledge that internal perturbations during the planetesimal-driven migration of the giant planets, such as planet-planet resonant encounters and scatterings \citep{Tsiganis2005,Morbidelli+2009} could have also played a role in establishing the secular architecture of the giant planets. Considering the robust statistics in the substellar flyby scenario found in our study, a combination of internal and external perturbations appears to be a likely part of the story of the solar system.

\section*{Data availability}
\noindent
The solar system ensemble used in this study can be found at \href{https://doi.org/10.5281/zenodo.4299102}{Zenodo.4299102} and the data generated for this article can be found at \href{https://doi.org/10.5281/zenodo.15288677}{Zenodo.15288677}.

\section*{Code availability}
\noindent
A repository containing a portion of the data underlying this article and code for running the simulations and generating the figures can be found at \href{https://github.com/zyrxvo/Secular-Origin}{github.com/zyrxvo/Secular-Origin}.
This research was made possible by the open-source projects 
\texttt{Jupyter} \citep{jupyter}, \texttt{iPython} \citep{ipython}, \texttt{matplotlib} \citep{matplotlib, matplotlib2}, \texttt{joblib} \citep{joblib} and \texttt{gnu-parallel} \citep{gnuparallel2024}.
The open-source projects \reb \citep{ReinLiu2012} and \airball \citep{airball} that enabled this research are also available.

\section*{Acknowledgements}
\noindent
G.B. is grateful to Mykhaylo Plotnykov, Dang Pham, and Sam Hadden for useful discussions.
The research of G.B. and H.R. has been supported by the Natural Sciences and Engineering Research Council (NSERC) Discovery Grants RGPIN-2014-04553 and RGPIN-2020-04513.
Their research was enabled in part by support provided by Digital Research Alliance of Canada (formerly Compute Canada; \href{https://alliancecan.ca/en}{alliancecan.ca}).
Computations were performed on the Niagara supercomputer \citep{SciNet2010, Ponce2019} at the SciNet HPC Consortium (\href{www.scinethpc.ca}{scinethpc.ca}). 
SciNet is funded by the following: the Canada Foundation for Innovation; the Government of Ontario; Ontario Research Fund -- Research Excellence; and the University of Toronto.
R.M. is grateful to the Canadian Institute for Theoretical Astrophysics (CITA) for hosting a sabbatical visit during which this project began. R.M. also acknowledges research support from NASA grant 80NSSC18K0397 and from the program “Alien Earths” (supported by the National Aeronautics and Space Administration under agreement No.80NSSC21K0593) for NASA’s Nexus for Exoplanet System Science (NExSS) research coordination network sponsored by NASA’s Science Mission Directorate. 
The citations in this paper have made use of NASA’s Astrophysics Data System Bibliographic Services.

\setcounter{section}{0}
\renewcommand{\thesection}{\Alph{section}}
\setcounter{figure}{0}
\renewcommand\thefigure{A.\arabic{figure}}   
\setcounter{table}{0}
\renewcommand\thetable{A.\arabic{table}}
\section{Appendix}
\label{sec:appendix}
\subsection{Likelihood of Multiple Flybys}
We can model the likelihood of the solar system experiencing a specific number of flyby encounters using the Poisson distribution.
This expresses the probability of a given number of events in a fixed time interval. 
We can use the Poisson distribution because the number of flyby events does not depend on any previous flyby events and we model the encounter rate using a constant (static) object density, velocity dispersion, and cross-section.
Given the encounter rate of $\Gamma \approx 1.3\cdot 10^{-4}\,  \mathrm{Myr}^{-1}$ (see \eq{eq:encounter-rate}), the probability of $k$ flyby events in time interval $\Delta t$ is given by ($k \in \mathbb{N}, t \in \mathbb{R}^{+}$)
\begin{equation}
    P(k,\Delta t) = \frac{(\Gamma \Delta t)^{k} e^{-\Gamma \Delta t}}{k!}\,.
\end{equation}
Therefore, the likelihood of zero flybys in $100\,\mathrm{Myr}$ is $P(0,100\,\mathrm{Myr}) \approx 0.9871$. 
Similarly, $P(1, 100\,\mathrm{Myr}) \approx 0.0128$, and so on.
We can express the probability of having \emph{at least} $k$ flybys as 
\begin{equation}
    \hat{P}(k \geq n,\Delta t) = 1 - \sum_{k=0}^{n-1} P(k,\Delta t)\,,
\end{equation}
such that the likelihood of the solar system experiencing \emph{at least} one flyby is $\hat{P}(k \geq 1, 100\,\mathrm{Myr}) \approx 0.0129$.

\subsection{Success Criteria}
\label{sec:success-criteria}
We discuss here the success rate for the four criteria defined in Section 3 of \citep{Nesvorny2012} resulting from the flyby simulations of this study. 
Criterion A is that four planets remain in the simulation.
Criterion B has three parts: all of the semi-major axes are within $20\%$ of the current solar system, the average eccentricities are all less than 0.11, and the average inclinations are less than 2$^\circ$.
See Table~\ref{tab:ecc-inc} for the average eccentricity and inclination values for the best matching flyby case.
Criterion C, discussed in this study, requires the amplitude of the $g_5$ mode in Jupiter's orbit to be greater than $0.022$, greater than half of the observed value.
Criterion D is not applicable to our study because the giant planets do not migrate in our simulations. Instead, a specific study of the effect of each flyby on the terrestrial planets should be considered.
The criteria are dependent; C can only be successful if B is successful, and B can only be successful if A is successful.
Using these criteria with our data we find success rates of A (41.33\%), B (19.33\%), and C (1.63\%).
These success rates also show that $1\%$ of flyby simulations result in a `close match' with the solar system.
If we consider an intersection of successful matches from these criteria with $\mathcal{M}$ the success rate is $0.23\%$.

\begin{table}[h]
   \centering
   \vspace{1em}
   \caption{
   Averaged over 20 Myr, we show a comparison of the eccentricities and inclinations (with respect to the invariant plane) of the giant planets (SS) and our best matching flyby system described in section~\ref{sec:bestmatch}.
   }
   \begin{tabular}{c|cc|cc} 
      \toprule
      \midrule
      Planet & \multicolumn{2}{c|}{Eccentricity} & \multicolumn{2}{c}{Inclination [$^\circ$]}\\
      $\,$ & SS & Flyby & SS & Flyby \\
      \midrule
      Jupiter & $0.0456$ & $0.0289$ & $0.366^\circ$ & $0.164^\circ$ \\
      Saturn & $0.0536$ & $0.0259$ & $0.902^\circ$ & $0.402^\circ$ \\
      Uranus & $0.0435$ & $0.0593$ & $1.022^\circ$ & $0.920^\circ$ \\
      Neptune & $0.0096$ & $0.0352$ & $0.669^\circ$ & $0.146^\circ$ \\
      \bottomrule
   \end{tabular}
   \label{tab:ecc-inc}
\end{table}

\needspace{2 \baselineskip}
\subsection{Angular Momentum Deficit}
\label{sec:AMD}
\begin{figure}[h]
    \centering
    \includegraphics[width=\columnwidth]{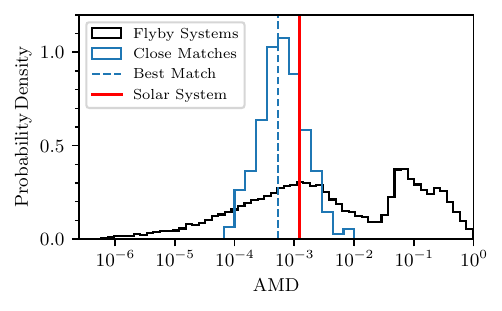} 
    \caption{Here we show in black the distribution of the normalized angular momentum deficit (AMD) compared to the solar system (shown in red).
    We highlight in blue the 422 flyby systems that are a \emph{close match} to the solar system according to our metric $\mathcal{M}$ along with the best matching system.
    We note that of the $5\cdot 10^4$ flyby systems, $15737$ have AMD within the range of our \emph{closely matching} systems; $6679$ are within the range of the best matching system; and $567$ are within $1\%$ of the solar system.
    \label{fig:amd}
    }
\end{figure}

We noted in Section~\ref{sec:metric} that some previous studies used the angular momentum deficit (AMD) to compare the orbital architectures of planetary systems.
However, we found that, for the purpose of comparing the similarity between flyby systems and the solar system, the AMD is not a useful parameter because the AMD can be dominated by a single planet which is a common result of flyby encounters.
There are many more flyby systems with AMD similar to the solar system than there are \emph{closely matching} flyby systems, see \fig{fig:amd}.
Flyby encounters tend to increase the eccentricities of the outermost planets more than the innermost planets \citep{Heggie1996}.
The results from our simulations are generally consistent with this theoretical expectation. 
Note that this is opposite to the solar system where Neptune has the smallest eccentricity out of the giant planets.
Approximately $8\%$ of the \emph{closely matching} cases have the same trend as in the solar system.

\subsection{Orbital elements}
\label{sec:oe}

\begin{figure}[t]
    \centering
    \includegraphics[width=\columnwidth]{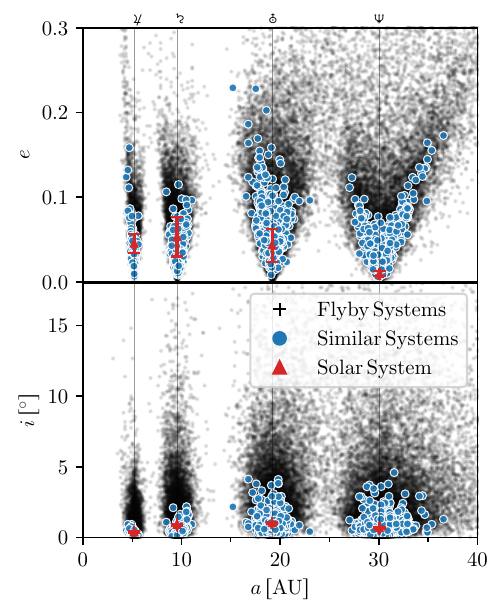} 
    \caption{
    The distributions of the osculating eccentricity, inclination, and semi-major axis of the outer solar system planets in each flyby simulation (black crosses) as well as the subset of those that led to a \emph{close match} with the solar system (blue dots, see also Fig.~\ref{fig:flyby-params}).
    For the solar system, we plot the average and standard deviation of the initial osculating orbital elements of the reference set of solar systems described in Section~\ref{sec:referencesystems}.
    \label{fig:kepler-params-all-oja}
    }
\end{figure}

We have argued in Section~\ref{sec:metric} that our new metric is well suited for finding close matches of flyby systems to the solar system. 
Here we show how this metric compares to the  approach of matching orbital elements adopted in some previous studies. 
The semi-major axis, eccentricity, and inclination of the solar system and the flyby simulations are shown in Fig.~\ref{fig:kepler-params-all-oja}.
We highlight (in blue dots) those flyby simulations that we have labeled a \emph{close match} using our log-spectral distance metric.
For the solar system, we use the ensemble of 20~Myr snapshots from the 5~Gyr solar system integrations described in Sec.~\ref{sec:referencesystems}.
Note that the orbital elements of the solar system, especially the eccentricities, vary significantly (indicated by their standard deviation over the solar system reference set).

We can see that, compared to the solar system, there is a larger range of orbital parameters in the flyby outcomes. 
The \emph{close match} cases do overlap with the solar system in the space of orbital parameters. 
For Jupiter, Saturn, and Uranus, the median eccentricities and inclinations of the \emph{close match} cases are comparable to those of the actual solar system. 
Neptune has a higher eccentricity in many (but not all) flyby simulations compared to the actual solar system because it is the outermost planet and thus most affected by distant flybys which tend to be the most common ones.

We stress that the reverse is not true: a good match in the space of orbital elements is not necessarily a good match when evaluated using our log-spectral metric.
To illustrate this point, we find the closest match by only comparing orbital elements of the flyby simulations to the orbital elements of the solar system (semi-major axis, eccentricity, and inclination). 
We do this using a least squares approach to define the following metric to compare the orbital elements:
\begin{eqnarray}
\mathcal{M}_{\rm oe} &=& \sum_{k=5}^8 \left(\frac{e_k-\bar{e}_k}{\bar{e}_k}\right)^2 + \sum_{k=5}^8 \left(\frac{i_k-\bar{i}_k}{\bar{i}_k}\right)^2 \nonumber\\
&& +0.1\sum_{i=5}^8 \left(\frac{a_i-\bar{a}_i}{\bar{a}_i}\right)^2 ,
\label{eq:oe-metric}
\end{eqnarray}
where the sum is over the four outer planets and the barred parameters are those of the present day outer solar system planets. 
Note that this metric is by no means unique and in fact quite arbitrary. For example, we added a relative weight of 0.1 to the fractional semi-major axis to avoid overfitting this component.

\begin{figure*}
    \centering
    \includegraphics[width=0.88\textwidth]{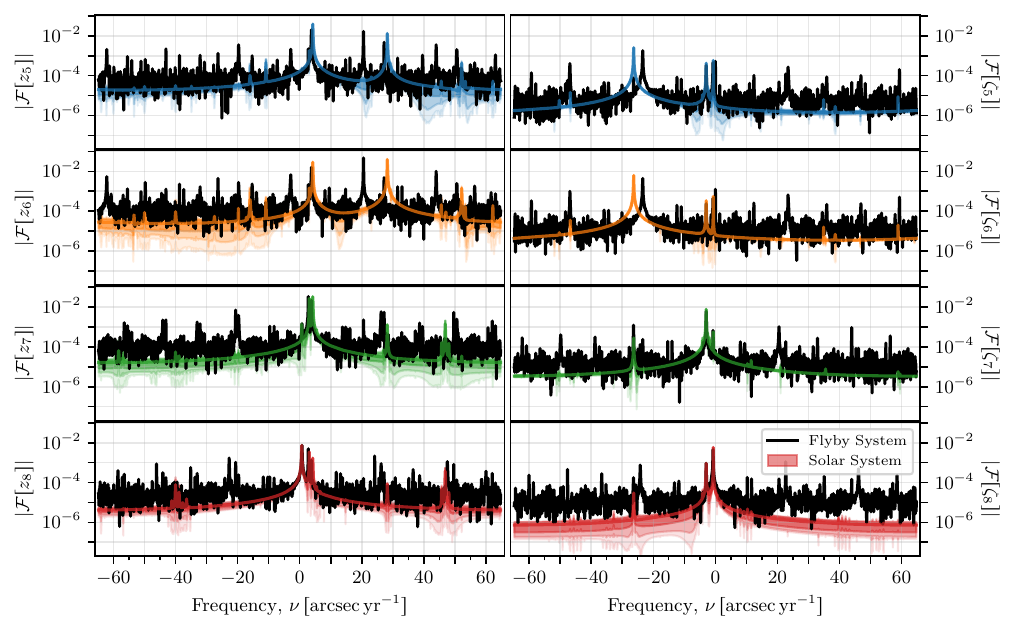}
    \caption{
    Power spectra of the complex eccentricities (left column) and complex inclinations (right column). 
    The black line shows the best matching flyby case found by comparing only orbital parameters (see \eq{eq:oe-metric}). 
    The color-shaded plots show the ensemble of spectra of the solar system integrations. 
    Only a subset of the peaks of the flyby system's power spectra match those of the solar system. 
    Aside from those matching peaks, the spectra are qualitatively very different from each other. 
    Our log-spectral metric is able to find much better matches as shown in Fig.~\ref{fig:power-spectra}.
    \label{fig:seems-close}
    }
\end{figure*}

We plot the spectrum of the closest match that we find using the $\mathcal{M}_{\rm oe}$ metric in Fig.~\ref{fig:seems-close}.
The flyby parameters for this encounter are $m_\star = 1.67\,\mathrm{M}_\mathrm{J}$, $q_\star = 4.7\,\mathrm{AU}$, $v_\infty = 2.4\,\mathrm{km\,s}^{-1}$, and $i_\star = 52^{\circ}$.
One can see that both the location and the amplitude of the peaks are indeed a good match to the solar system since they are determined by the orbital parameters we compare. 
However, the spectrum looks otherwise quite different than that of the solar system. 
There is a wide range of other frequencies with amplitudes that are several orders of magnitude higher than those of the solar system. 
One can also see frequencies and their harmonics which are not present in the solar system.
This example demonstrates the need to go beyond comparing orbital parameters or comparing only the fundamental secular modes. 
We think that our new log-spectral distance metric can more reliably identify a simulated system that matches the secular architecture of the solar system. 

\subsection{Null Hypothesis}
\label{sec:null}

\begin{figure*}
    \centering
    \includegraphics[width=0.88\textwidth]{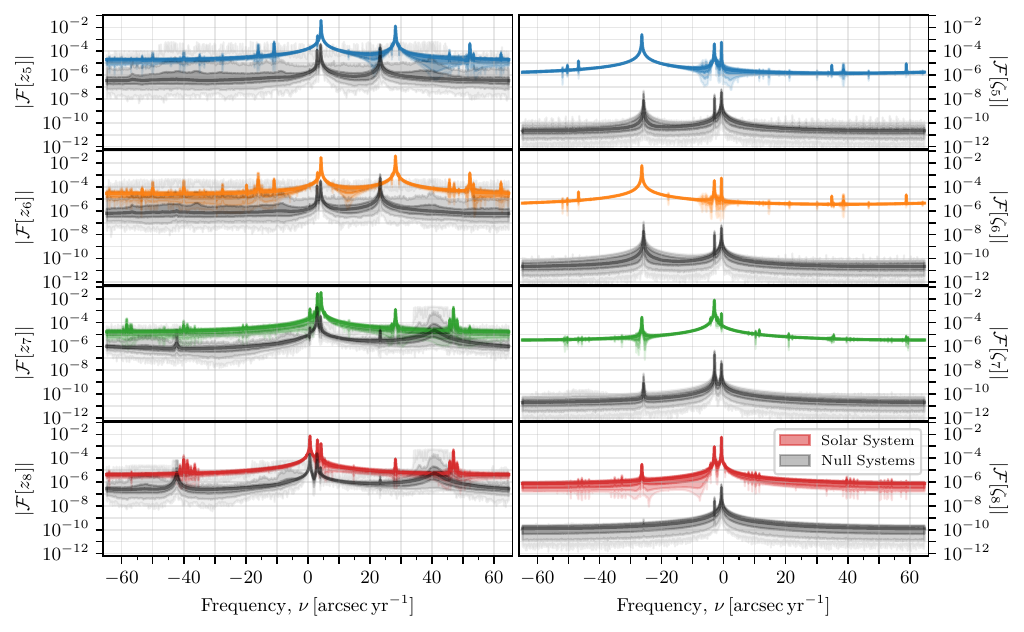}
    \caption{
    Testing the "null hypothesis": the power spectra of an ensemble of 1000 integrations of the giant planets initialized in circular and very nearly coplanar orbits, integrated for 20 Myr, with no stellar flybys.
    The median power at each frequency of the ensemble is shown as a black line, with gray-shaded regions representing the one–, two– and three–σ deviations. 
    For comparison, the color-shaded plots are the actual solar system's power spectra.
    This shows that, from initially very nearly circular and coplanar orbits (at their observed semi-major axes), planet-planet interactions alone cannot account for the observed amplitudes of the giant planets' secular modes. 
    See Section~\ref{sec:null} for details.
    \label{fig:null}
    }
\end{figure*}

Here we test the ``null hypothesis", namely, given our circular, coplanar initial conditions, can the effect of planet-planet interactions alone (without any flybys) generate the secular mode amplitudes found in the observed solar system?
We start with a random subset of $1000$ configurations from the same initial conditions we used in our set of $5\cdot 10^{4}$ flyby simulations.
However, because a \emph{perfectly} coplanar, point-mass planetary system simulation will never generate \emph{any} relative inclinations we slightly perturb the planets' initial vertical positions and velocities on the order of $10^{-8}$ fraction of their orbital radius and orbital velocity, respectively.
As with other simulations throughout this work, we integrate this set for $20\,\mathrm{Myr}$ taking $2^{11}$ evenly-spaced samples.
The power spectra of the resulting complex eccentricities and complex inclinations in this ensemble are shown in \fig{fig:null}. 
For comparison, the power spectra of the actual solar system today are also plotted in this figure.

We briefly comment on these results.
As expected (and noted in Section~\ref{sec:results}), the choice to initialize the giant planets on their current semi-major axes results in fundamental secular mode frequencies similar to those observed in the solar system today. 
We see this in \fig{fig:null} for both the eccentricity and inclination power spectra.
However, the maximum amplitudes of the eccentricities are on the order of $10^{-3}$ (roughly $\sim\mathrm{M_J}/\mathrm{M}_\odot$),
and the inclination amplitudes are all below about $10^{-6}$; the latter underscore the low efficiency of inclination excitation from a coplanar initial state.
Comparing with the power spectra of the actual solar system, it is evident that the planet-planet interactions of the giant planets alone cannot excite the eccentricity and relative inclination amplitudes observed in the solar system today.

\begin{figure*}
    \centering
    \includegraphics[width=\textwidth]{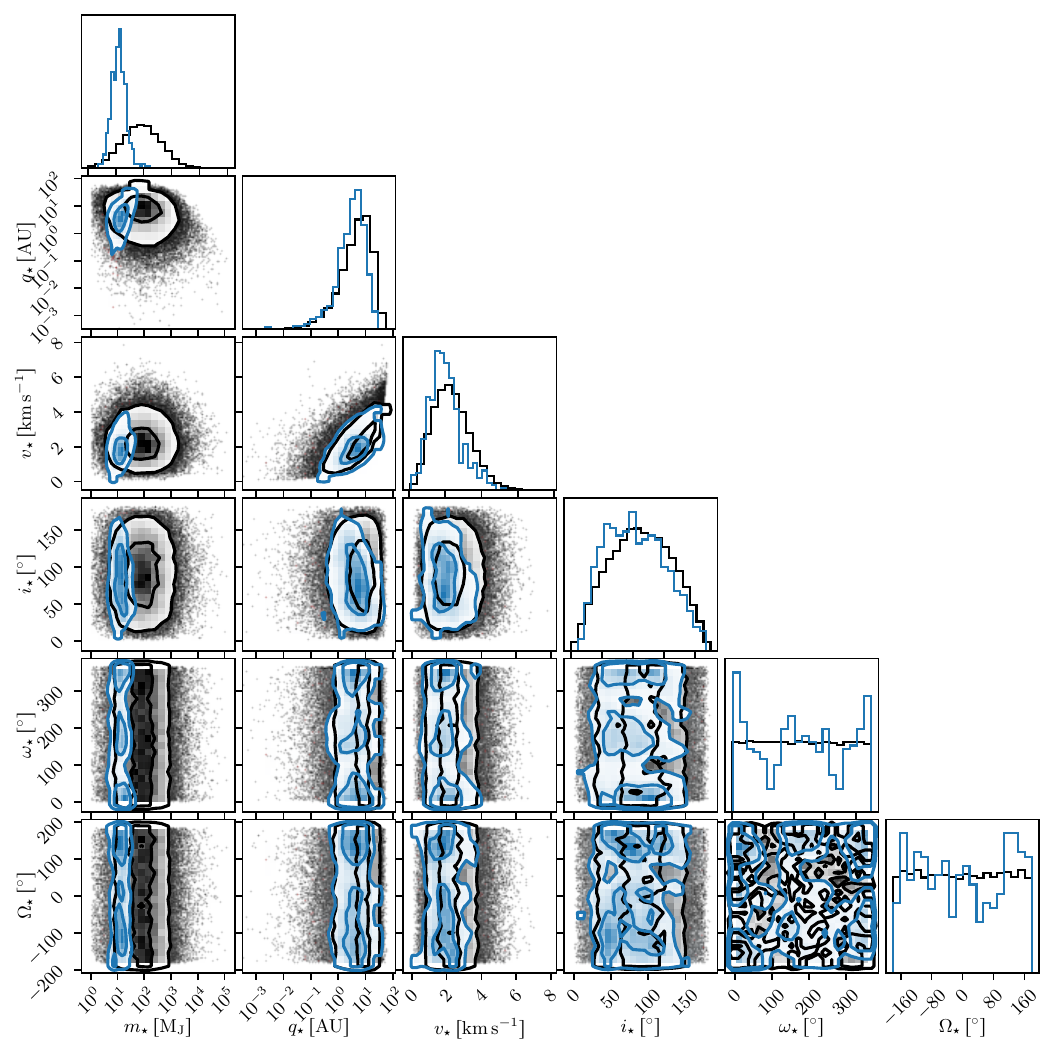}
    \caption{Here we show the distributions of all the flyby parameters in our simulations (black) as well as the subset of those parameters that led to a \emph{close match} with the solar system (blue).
    The contours show the first two standard deviations.
    See Section~\ref{sec:results} for more details.
    \label{fig:all-flyby-params}
    }
\end{figure*}
\newpage
\bibliography{full}

\end{document}